# Macroscopic quantum effects of electromagnetic induction in silicon nanostructures

L.E. Klyachkin✉, N.T. Bagraev, A.M. Malyarenko

Ioffe Physical Technical Institute of the Russian Academy of Sciences, 26, Polytekhnicheskaya St., St. Petersburg, 194021, Russia

✉ leonid.klyachkin@gmail.com

**Abstract.** At room temperature, a macroscopic quantum galvanomagnetic effect of Faraday electromagnetic induction was demonstrated under conditions of the capture of single magnetic flux quanta in the edge channels, confined by chains of negative-U centers, in a silicon nanostructure heavily doped with boron, prepared in Hall geometry on an n-type Si (100) substrate. It is shown that this effect leads to the appearance of an induction current when only a constant magnetic field is applied in the absence of an externally applied voltage or a stabilized current. The experimental dependences of $U_{xx}$, $U_{xy}$, and $U_{pn}$ on the magnitude of the external magnetic field in its various directions demonstrate both the Hall staircase of conductivity and the Shubnikov–de Haas oscillations.

*Acknowledgements.* The work was financed within the framework of the State task on the topic 0040-2019-0017 "Interatomic and atomic-molecular interactions in gases and condensed media; quantum magnetometry and multiphoton laser spectroscopy".

**Citation:** Klyachkin LE, Bagraev NT, Malyarenko AM. Macroscopic quantum effects of electromagnetic induction in silicon nanostructures. *Materials Physics and Mechanics*. 2022;xx(x): xxx-xxx. DOI: 10.18149/MPM.xxx2022_x.

## 1. Introduction

In recent years, the creation of various types of nanostructures has led to the intensive development of nanoelectronics. During research in this area, a number of macroscopic quantum phenomena were discovered, which manifest themselves at high temperatures up to room temperature. Such phenomena include high-temperature oscillations of Shubnikov–de Haas, de Haas–van Alphen, Aharonov–Bohm, Hall quantum resistance staircase, longitudinal conductivity quantum ladder, and Faraday electromagnetic induction under conditions of the capture of single magnetic flux quanta. Such effects have been observed in graphene, as well as in similar topological insulators and superconductors [1-3]. In addition, high-temperature quantum phenomena have been detected in silicon nanostructures (SNSs) based on single-crystal silicon, 6H-SiC, and $CdF_2$ containing quantum wells (QWs) with edge channels formed from chains of centers with a negative correlation energy (negative-U). It was experimentally found that these nanostructures exhibit the properties of topological insulators [4-7]. It should be noted that negative-U centers can have a different nature: in particular, in single-crystal silicon and CdF$_2$ based SNSs they are formed on the basis of B$^+$–B$^-$ boron dipoles oriented in the (111) direction under conditions of heavy doping, while in 6H-SiC based nanostructures negative-U centers have a vacancy nature [8]. It was shown that under compression conditions, due to high pressure of the order of hundreds of GPa [9], in such nanostructures, an edge channel with a cross-section of 2×2 nm appears on the surface of the QW limited by negative-U centers, which leads to an increase in the carrier relaxation time [10], due to the suppression of electron-electron interaction (EEI). Note that in a number of works, the neutralization of EEI in edge channels was predicted in cases when they are limited by chains of d- or f-elements [11,12]. Thus, it was experimentally demonstrated that layers consisting of negative-U dipole centers that form edge channels on the QW surface determine the possibility of observing macroscopic quantum phenomena at high temperatures up to room temperature [4-7].

Among the observed macroscopic quantum phenomena in the SNSs, one should especially highlight the Faraday electromagnetic induction (FEMI) under the conditions of the capture of single magnetic flux quanta [13]. It was found [7,14] that during the longitudinal current flow in a system with an edge channel, single magnetic flux quanta are captured by single charge carriers in the edge channel and the effect of electromagnetic induction is realized, which manifests itself in electrical measurements, and also leads to intense radiation in far IR, THz and GHz wavelengths. Taking into account the results of the above experiments, one should expect the manifestation of a galvanomagnetic effect in such nanostructures, namely, the appearance of an induction current when only a constant magnetic field is applied in the absence of an externally applied voltage or the transmission of a stabilized current. In this case, the capture of single quanta of the magnetic flux on single carriers in the edge channels will be determined only by varying the magnitude and direction of the magnetic field. It is expected that in this case, the effect will manifest itself when studying the dependences of $U_{xx}$, $U_{xy}$, and $U_{pn}$ on the magnitude of the external magnetic field in a nanostructure formed in the Hall geometry.

## 2. Experimental methodology

SNS is an ultra-narrow p-type silicon quantum well bounded by δ-barriers heavily doped with boron (5×10$^{21}$ cm$^{-3}$) on the surface of n-silicon (100) (Fig. 1). Boron atoms in δ-barriers form trigonal dipole centers with negative correlation energy (B$^+$–B$^-$) due to the negative-U reaction: $2B_0 \rightarrow B^+ + B^-$ [5] limiting edge channels. Thanks to this circumstance, the EEI is significantly suppressed in such an edge channel, as a result of which the carriers have a long relaxation time, which, in turn, makes it possible to observe macroscopic quantum processes at high temperatures up to room temperature.

The studies of the de Haas–van Alphen (dHvA) [4] and Shubnikov–de Haas (SdH) effects, as well as the quantum Hall effect (QHE) [5], showed that in these experimental SNSs, the two-dimensional concentration $p_{2D}$ of single charge carriers in the edge channel bound by chains of negative-U centers is 3×10$^{13}$ m$^{-2}$. Their behavior is characterized by a long relaxation time and a low value of the effective mass [15,16]. When studying dHvA in these experimental SNSs, it was shown that as the magnetic field increases, the static magnetic susceptibility begins to reveal dHvA oscillations due to the creation of Landau levels, $E_\nu = \hbar\omega_c(\nu+1/2)$, $\nu$ is the number of the Landau level. In particular, the condition of covering the edge channel with single magnetic flux quanta turns out to be satisfied at an external magnetic field strength $\Delta B$ = 124.6 mT, which follows from the relation $\Phi_0 = \Delta B S$ and corresponds to the filling of the first Landau level, $\nu_1 = 1$, $\nu_1 = p_{2D}h/eB$, where $\Phi_0 = h/e$ (or $h/2e$) is the magnetic flux quantum, and $S$ is the area on which the capture is made. Based on the foregoing, it is possible to estimate the longitudinal size $l_0$ of the region of the edge channel, in which the interference of a single carrier occurs:

$$l_0 = \frac{\Phi_0}{\Delta B d_0}, \qquad (1)$$

where $\Delta B$ = 124.6 mT; $d_0$ = 2 nm is the width of the edge channel [5] since this value is the characteristic distance between the barriers containing negative-U centers that form the edge channels bounding the QW. From here, we get the value $l_0 \approx 16$ μm. It can be said with confidence that the edge channel of the investigated nanostructure consists of regions of interference of single carriers, taking into account the value of their two-dimensional density, determined from Hall measurements, $3 \times 10^{13}$ m$^{-2}$. This value of the two-dimensional density corresponds to the distance between carriers (holes) in the edge channel of approximately 16 μm. In other words, each region of the edge channel, in which a single carrier ('pixel') is located, consists of layers containing boron dipoles with an area of $S_{pixel}$ = 16 μm × 2 nm, along which the carrier tunnels. The quantum well is limited by these two layers with a width and height of approximately 2 nm, which is consistent with the depth of the diffusion profile [17].

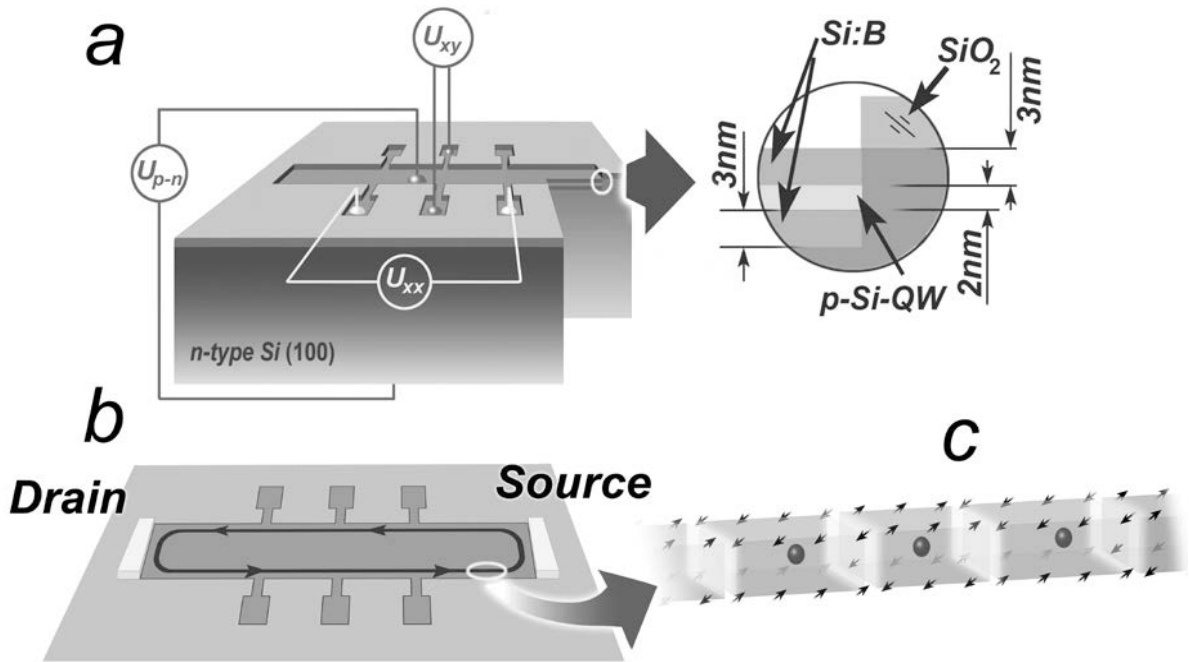

**Fig. 1.** Silicon nanostructure (a) containing a heavily doped silicon quantum well bounded by δ-barriers consisting of negative-U dipole boron centers and forming an edge channel (b) with ballistic conductivity in it. Fragment of the edge channel (c), consisting of a sequence of regions of interference of single carriers (pixels)

Thus, due to the FEMI, the quantum interference of single carriers occurs inside the pixels of the edge channel, and, in accordance with the classical FEMI relation, it is possible to estimate the induction current $I_{ind}$ arising due to the application of an external magnetic field:

$$I_{ind} = \frac{\Delta E}{\Delta \Phi} = \frac{neV_g}{m\Phi_0}, \qquad (2)$$

where $\Delta E$ is the change in energy when varying the magnitude of the magnetic flux in the region of quantum interference; $\Delta \Phi = \Delta B S$; $\Delta B$ is the change in the external magnetic field inside the region of quantum interference with area $S$; $n$ is the number of single charge carriers in the region of quantum interference; $m$ is the number of magnetic flux quanta captured in the region of quantum interference; $V_g$ is the control voltage applied to the edge channel [16]. In particular, if we consider quantum interference and $I_{ind}$ in a single pixel containing a charge carrier, then $n = 1$, $m$ is the number of magnetic flux quanta captured by a single pixel, and the resulting conductivity value in $e^2/h$ units is greater than one. At the same time, under conditions of quantum interference, fractional conductivity values with $\nu > 1$ ($\nu = n/m$) can appear on an area larger than the area of one pixel [5]. With an increase in the magnetic field, quantum interference occurs exclusively in single pixels of small size. In this case, one should expect the capture of several magnetic flux quanta per pixel, and, accordingly, fractional conductivity values with $\nu < 1$, which manifests itself when registering the Hall staircase of conductivity [5].

Hence, it follows that the FEMI can play an important role in the observation of macroscopic quantum effects in SNSs. In addition, it becomes possible to implement device structures based on the application of a control voltage to edge channels, an example of which is a spin transistor [18]. This device can operate under the conditions of transport of single carriers when registering both longitudinal and transverse voltage drops. Such an effect can only be obtained by applying an external magnetic field, which is the cause of the induction current. It should be noted that the proposed version of the FEMI quantum effect can occur even in the absence of a specified source-drain current in the Hall geometry since a change in the external magnetic field is sufficient to generate $I_{ind}$. Thus, macroscopic quantum phenomena can be detected using galvanomagnetic techniques.

The idea of the presence of the FEMI in quantum-well structures was first proposed by Laughlin [19], who suggested using a special wave function, within which the electron-electron interaction is suppressed, which makes it possible to explain the appearance of the fractional Hall conductivity staircase [20]. In this case, it was not required to divide the edge channel into pixels with forced localization of single carriers in them [19,21]. However, the methods used to register the fractional QHE revealed some contradiction with the condition of its observation in a system with a free gas of carriers, namely, for the experimental implementation of the fractional QHE, pre-illumination with monochromatic light at low temperatures were used, as well as the application of voltage to the gate during cooling of the samples [20,22]. Moreover, selective illumination satisfied the condition of recharging DX centers and other negative-U centers with their transition to the dipole state [5,22-26]. That is, the preliminary preparation of the sample for the registration of macroscopic quantum effects probably corresponded to the creation of conditions for suppressing the electron-electron interaction even at a high carrier density.

Thus, in accordance with expression (2), integer and fractional quantum phenomena can occur in experiments with SNSs under conditions of varying filling of pixels with different numbers of magnetic flux quanta. Using the scaling effect, i.e. by changing the distances between the contacts (using different contacts of the SNS made in the Hall geometry for the research), it is possible to change the number of carriers in the edge channel and obtain fractional values both greater and less than one [5].

It should be taken into account that in the investigated SNS on the area of the edge channel $S_{xx}$ between the XX contacts ($S_{xx}$ = 2 mm × 2 nm = $4\times10^{-12}$m$^2$) there are approximately 125 single carriers, which corresponds to the value of the two-dimensional density $p_{2D} \approx 3 \times 10^{13}$ m$^{-2}$. The results of measurements of the field dependences of the magnetic susceptibility [4] and Hall measurements [5] are in good agreement with this value.

The above-described SNSs containing edge channels consisting of pixels exhibit the property of quantized conductivity [4]. As a result, we can consider the edge channel as a ballistic one, in which each of the pixels is characterized by a resistance equal to the resistance quantum $h/e^2$. In addition, it was shown [7] that in such edge channels it is possible to form double-length pixels with a resistance $h/2e^2$ containing a pair of carriers with the probable implementation of a Josephson transition near the pixel boundary.

In the framework of this study, the detection of inductive current when only a magnetic field is applied is performed by measuring the voltage that occurs under the influence of a magnetic field on the contacts XX ($U_{xx}$), XY ($U_{xy}$), and on the gate contacts ($U_{pn}$). In this case, the induction current is determined from a simple expression:

$$U = I_{ind} R, \tag{3}$$

where $R$ is the resistance of the section of the edge channel between the measuring contacts, and $U$ is the voltage that occurs on these contacts.

The SNSs studied within this paper are made in the Hall geometry (Figure 1), so their edge channels, in which the induction current ($I_{ind}$) is induced, are located in the XX direction. Previously, it was found [5] that during the technological process of QW formation, the SNS surface is covered with a field of fractal-type pyramids with an apex angle of 55°, which are oriented in the (111) direction. These pyramids are an ordered accumulation of their own interstitial silicon atoms and are separated from each other by vacancies, which, in turn, are transformed in the process of doping into the above described trigonal negative-U dipole centers, also oriented in the (111) direction. Previously, it was shown [4] that the chains of such trigonal negative-U centers, which form δ-barriers that limit the edge channel, are oriented in the (011) direction.

In addition, studies of the dHvA effect [4] and spin-dependent transport [27] in SNSs showed that the characteristics of ballistic carrier transfer in the edge channel in the XX direction are identical to the XY direction.

It follows from the foregoing that it is necessary to conduct experimental studies of $U_{xx}$, $U_{xy}$, and $U_{pn}$ on the magnitude of the external magnetic field at different orientations of the magnetic field in a plane parallel to the XY direction. Therefore, four orientations of the magnetic field were chosen in these experiments: 1) Hall (the magnetic field is perpendicular to the SNS); 2) the magnetic field is parallel to the SNS and XY contacts; 3,4) intermediate cases with a deviation of the direction of the magnetic field from the Hall direction by 35° and 55°.

The magnetic field was sweeping in the range of 0–520 mT with a step of about 0.05 mT, then turned off, and then the measurements were repeated with the opposite polarity of the magnetic field. On the experimental dependences presented below, for the definite direction of the magnetic field, we showed both as $B^+$ (the magnetic field is directed towards the SNS) and $B^-$ (the magnetic field is directed away from the SNS).

The $U_{xx}$, $U_{xy}$ and $U_{pn}$ values were measured using KEITLEY 2182A and Agilent 34420A nanovoltmeters.

## 3. Experimental results

Since $I_{ind}$ arises in each pixel as a result of the capture of magnetic flux quanta, and the connection between pixels is carried out by tunneling single carriers through dipole centers in the barriers limiting them, parallel connection of pixels is realized in the edge channels. Thus, it can be expected that the value of the longitudinal registered voltage will correspond to the value $U_{ind} = I_{ind}(h/e^2) \cdot (1/k)$, where $k$ is the number of pixels containing single charge carriers between the measuring contacts, and the corresponding $I_{ind}$ is determined from Faraday relation (2).

The above consideration makes it possible to describe the presence of a conductivity (resistance) staircase that arises when the magnitude of the external magnetic field changes within the framework of the Hall geometry of the experiment. The presence of such a staircase is expected both when registering $U_{xy}$ in the perpendicular orientation of the external magnetic field and, accordingly, when registering $U_{xx}$ when the field is oriented along the SNS plane parallel to the XY contacts, and also when registering $U_{pn}$ (at the p-n junction on the gate-substrate contacts) in both of the above-mentioned field directions.

However, the contribution to the measured quantities can give the SdH effect. Then the role of the resistance in expression (3) will be determined by the Landau size quantization, which gives an oscillatory dependence of the resistance: if the Fermi level coincides with the Landau level, the resistance peaks, and if they do not coincide, the resistance cancels out [28]. In this case, characteristic oscillations are expected in the $U_{xx}$ dependences at a perpendicular orientation of the external magnetic field.

Note that the position of the Landau level peaks corresponds to the middle step of the Hall resistance staircase [28].

It follows from the above that it is expedient to use the interrelationship of $U_{xx}$ and $U_{xy}$, namely $R_{xx} = R_{xy}/k$, where $k$ is the number of pixels between the measurement contacts XX. Since $R_{xy}$ is determined by the resistance of a pixel containing a single carrier ($h/e^2$), the behavior of $R_{xx}$ with a change in the magnetic field follows the corresponding dependence of $R_{xy}$, and $U_{xx} = I_{ind}R_{xx} = I_{ind}R_{xy}/k$.

**Magnetic field perpendicular to the SNS.** Figure 2 shows the experimental dependences of $U_{xx}$, $U_{xy}$, and $U_{pn}$ on the magnitude of the external magnetic field oriented perpendicular to the SNS plane. $U_{xx}$ dependences (Fig. 2a) demonstrate pronounced SdH oscillations, which are determined, first of all, by the mechanism of successive capture of magnetic flux quanta on the pixels of the edge channel with successive generation of induced current. Moreover, on each of the presented dependences, several periods of oscillations are observed. Previous studies of the dHvA [4], SdH, and QHE [5] effects showed that the period of oscillations when a magnetic flux quantum is captured by a single pixel with an area $S_{pix} = 16.6\mu m \times 2$ nm = 33.2 $10^{-15}$ m$^2$ is $\Delta B$ = 124.6 mT. All curves shown in Fig. 2a demonstrate oscillations at 124.6 mT, and in addition, all of them have additional sets of oscillation periods, which are determined by more complex combinations of pixels, including those associated with SNS defects, which cannot be calculated.

$U_{xy}$ (Fig. 2b) and $U_{pn}$ (Fig. 2c) dependences demonstrate a stepwise character, which indicates the dominance of the QHE in comparison with the SdH effect for a given direction of the external magnetic field. However, along with the steps, these dependences also exhibit weak SdH oscillations, which most likely indicates the implementation of a mixed regime with the presence of both of the above effects. Note that the change in the polarity of the magnetic field in the study of the dependence $U_{xy} = f(B)$ leads to a pronounced Zeeman effect (Fig. 2b).

Since the SNS is only affected by the magnetic field (neither a bias voltage nor a stabilized current $I_{ds}$ are applied to it), we can estimate the current induced by the magnetic field, since it is known that a pixel, under the influence of a current, is a radiation generator at its resonant frequency of 2.6 THz [6]. The quantum energy $h\nu$ will be determined by the formula:

$$h\nu = I_{ind}\Delta\Phi, \tag{4}$$

where $\nu$ is the generation frequency, $\Delta\Phi$ is the change in the magnetic flux through the pixel. The frequency of generation can be predicted by knowing the dimensions of the resonator.

Taking into account the Wulff–Bragg relation $2nd = \lambda$, (where $d$ is the resonator length, i.e. the pixel length, $n$ is the refractive index, $\lambda$ is the radiation wavelength), $I_{ind}$ can be estimated using the formula:

$$I_{ind} = \frac{ch}{2dn\Delta\Phi}. \tag{5}$$

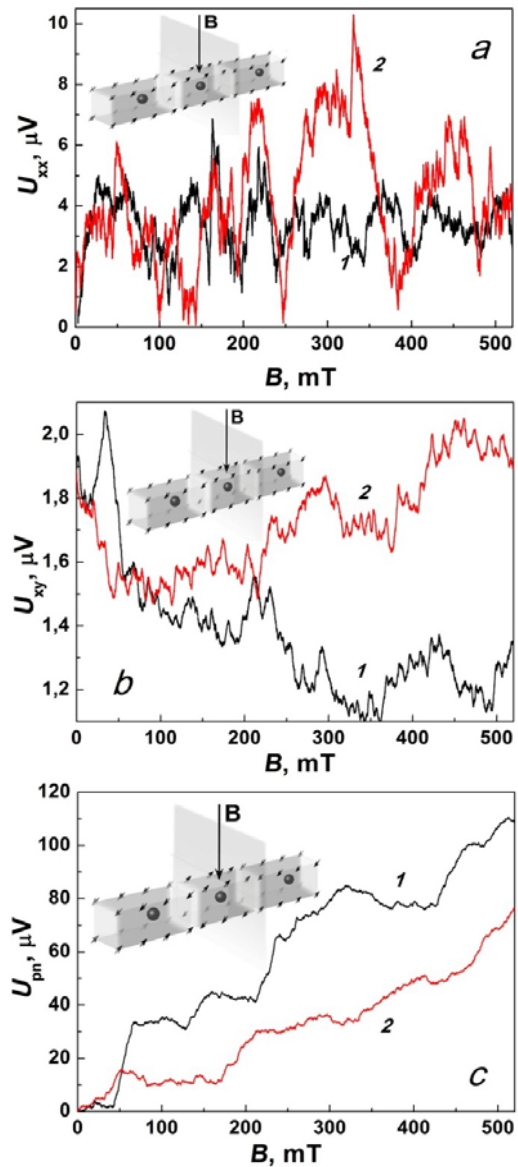

**Fig. 2.** $U_{xx}$ (*a*), $U_{xy}$ (*b*), and $U_{pn}$ (*c*) dependencies on the magnitude of the external magnetic field oriented perpendicular to the plane of the silicon nanostructure.
Dependencies: 1 – the field is directed to the surface of the SNS; 2 – the field is directed away from the SNS surface. T = 300 K

Then, knowing that the pixel length is 16.4 μm, assuming that the refractive index is close to the refractive index of silicon 3.4, and the change in the magnetic flux for the SNS at a magnetic field of 124 mT is equal to the magnetic flux quantum $\Phi_0 = h/e$ (or $h/2e$), we obtain $I_{ind}$ values of approximately 0.4 and 0.8 μA, respectively, for two values of $\Phi_0$.

It is necessary to take into account the contribution of two more components to the induced current: 1) the Josephson induced current $I_{indJ}$ [29,30] and 2) the superconducting induced current $I_{indS}$ arising due to the presence of the superconducting gap [13], which can be calculated by the formula:

$$I_{indJS} = e\Delta/h. \tag{6}$$

Note that in the case of the Josephson-induced current, a pair of single carriers must be considered, in which case the pixel size is doubled.

In this case, $I_{indJ}$ is approximately equal to 0.2 μA, considering that $\nu = 2.6$ THz and $I_{indS}$ is approximately equal to 0.8 μA, considering that $\Delta$ the width of the superconducting gap, is equal to 22 meV.

Now, knowing $I_{ind}$, we can obtain the calculated values of $U_{xy}$ using the formulas:

$$U_{xy} = 2I_{ind}\left(\frac{h}{e^2}\right) \tag{7a}$$

or

$$U_{xy} = 2I_{ind}\left(\frac{h}{2e^2}\right). \tag{7b}$$

Taking into account all the components of the inductive current, and also that their superposition can both be added up and subtracted, we obtain the calculated $U_{xy}$ values of the same order as the experimental ones, which are shown in Fig. 2b.

**The magnetic field in the SNS plane parallel to the XY contacts.** Figure 3 shows the experimental dependences of $U_{xx}$, $U_{xy}$, and $U_{pn}$ on the magnitude of the external magnetic field oriented in the SNS plane parallel to the XY contacts. In this case, the size and orientation of the edge channel pixels are preserved. However, compared to the previous orientation of the magnetic field, the carriers rotate along the field in a transverse plane relative to the SNS plane. In this case, additional compression of the carrier trajectory from below by the p-n junction field occurs, as a result of which the spread of parameters is smaller compared to the previous case.

$U_{xx}$ (Fig. 3a), $U_{xy}$ (Fig. 3b), and $U_{pn}$ (Fig. 3c) dependences, in the case of application of an external magnetic field in the SNS plane parallel to the XY contacts, demonstrate a stepwise character, which indicates a clear dominance of the QHE.

Note that the experimental values of $U_{xx}$ and $U_{pn}$ are 1-2 orders of magnitude higher than the values of $U_{xy}$, although when measuring the classical and quantum Hall effect, namely when applying a longitudinal stabilized current $I_{ds}$ to the SNS the measured voltage $U_{xx}$ is always significantly lower (by several orders of magnitude) than $U_{xy}$. In these experiments, when $I_{ds}=0$, $U_{xx}$ is approximately 2 orders of magnitude higher than the $U_{xy}$ values measured under the same conditions. This is explained by the appearance of strong polarization due to the Zeeman effect, which contributes to the $U_{xx}$ values by several orders of magnitude greater than the magnetic field-induced voltage $U_{xx\,ind}$ since the Zeeman polarization coefficient depends exponentially on the magnetic field. The same circumstance also explains the monotonic growth of the experimental dependences of $U_{xx}$ on the magnetic field strength.

Taking this into account, it is possible to estimate the magnitude of the voltage $U_{xxind}$ induced by the magnetic field using the following formula:

$$U_{xxind} = U_{xx} \exp\left(-\frac{2eB\hbar}{kTm^*}\right), \qquad (8)$$

where $e$ is the electron charge, $B$ is the magnitude of the magnetic field, $k$ is the Boltzmann constant, $T$ is the temperature, $m^*$ is the effective mass of charge carriers. If $m^*$ is assumed to be approximately $10^{-34}$ kg (previously it was shown in [15] that in SNS it is about $10^{-3} m_0$ at room temperature), then with a magnetic field $B = 124.6$ mT and at room temperature, the exponent will be $4.5\times10^{-5}$ and the value of $U_{xxind}$ will be about 10 nV.

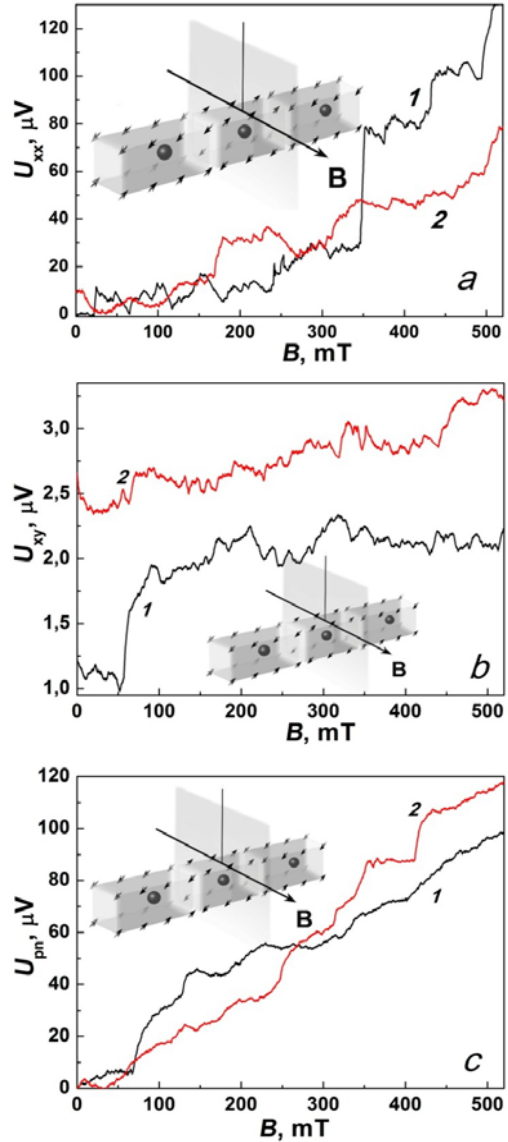

**Fig. 3.** $U_{xx}$ (*a*), $U_{xy}$ (*b*), and $U_{pn}$ (*c*) dependencies on the magnitude of the external magnetic field oriented along the plane of the silicon nanostructure parallel to the XY contacts. The field for dependencies (1) is directly opposite to the field for dependencies (2). T = 300 K

Taking into account the pronounced stepwise nature of the experimental dependences, it is possible to estimate the coincidence of the position of the steps in the $U_{xx} = f(B_{xy})$ and $U_{pn} = f(B_{xy})$ with integer and basic fractional values of the pixel resistance $R_{xx}$ calculated in $h/e^2$ units according to the formula:

$$R_{xx} = \frac{\Phi}{ne} = \frac{BS}{ne}[Ohm] = \frac{BS}{ne}\frac{e^2}{h}\left[\frac{h}{e^2}\right],\tag{9}$$

where $n = 1$; $S = 33.2 \times 10^{-15} m^2$.

The results presented in Fig. 4 show good agreement.

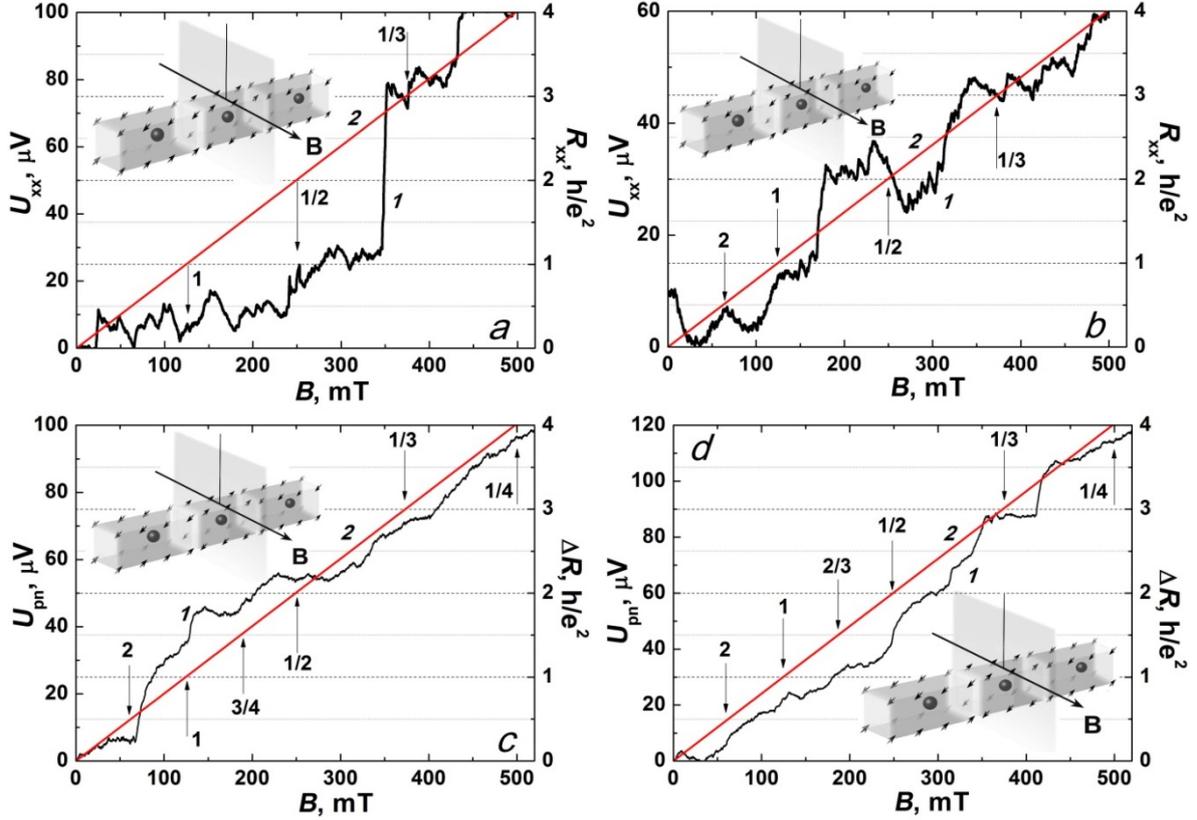

**Fig. 4.** Comparison of the position of the steps in the dependences (*a*, *b*, curves *1*) $U_{xx} = f(B_{xy})$ and (*c*, *d*, curves *1*) $U_{pn} = f(B_{xy})$ with integer and basic fractional calculated values of the pixel resistance $R_{xx}$ in $h/e^2$ units (straight lines *2*). The fields (a) and (b), (c), and (d), respectively, have the opposite direction. T = 300 K

**The magnetic field in intermediate positions between the perpendicular to the SNS plane and parallel XY contacts**. In this section, we consider cases of intermediate angles of the direction of the magnetic field relative to the SNS surface with a deviation of the direction of the magnetic field from the Hall direction by 35° and 55° in the plane perpendicular to the surface of the SNS and parallel to the XY contacts. Figure 5 shows the $U_{xx} = f(B)$ dependences, which have a mixed oscillatory-stepwise character.

We will consider the magnitude of such a deflected magnetic field as a superposition of two components: the Hall field ($B_{Hall}$) and the field directed along the SNS surface parallel to the XY contacts ($B_{H \| XY}$). It is obvious that the component of the magnetic field $B_{Hall}$ determines the oscillating nature corresponding to the SdH effect, while the other component of the magnetic field $B_{H\|XY}$ determines the stepwise nature of the $U_{xx} = f(B)$ dependences as a result of the dominance of the QHE. In Figure 5, the values of both components of the magnetic field are presented on additional abscissa axes. It can be seen that in the case when the $B_{Hall}$ component increases faster than $B_{H\|XY}$ with the increasing magnetic field (Fig. 5a), the $U_{xx}$ dependences are stepwise due to the dominance of the QHE. In the opposite case, when the $B_{H\|XY}$ component increases faster than $B_{Hall}$ with increasing magnetic field (Fig. 5b), at low fields the SdH effect dominates, and at higher fields a mixed regime occurs, in which both SdH and QHE effects are manifested.

In addition, in Figure 5 (line *3*), the positions of the pixel resistance $R_{xx}$ are indicated in $h/e^2$ units with integer and basic fractional calculated values depending on the value of $B_{H\|XY}$. The results obtained correlate well with the experimental dependences described above for various variants of applying a magnetic field in the cases considered above.

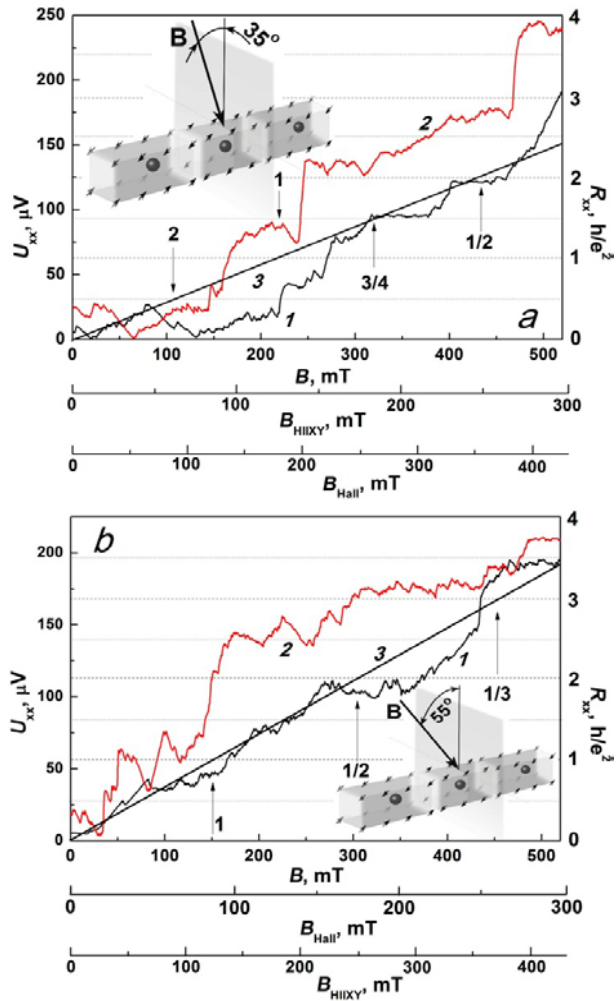

**Fig. 5.** Comparison of the position of the steps in the $U_{xx} = f(B)$ dependences (curves *1* and *2*) obtained with a deviation of the magnetic field direction in the plane perpendicular to the SNS surface and parallel to the XY contacts from the Hall direction by (a) 35° and (b) 55°, with integer and basic fractional calculated values of the pixel resistance $R_{xx}$ in $h/e^2$ units (straight lines *3*). For dependencies (*1*) and (*2*), the direction of the field is directly opposite. $T = 300$ K

## 4. Conclusions

The suppression of the electron-electron interaction, which leads to an increase in the relaxation time due to the presence of negative-U centers limiting the QW edge channels, makes it possible to detect macroscopic quantum effects, the Hall staircase of conductivity, and SdH oscillations at high temperatures up to room temperature. These effects arise due to FEMI under conditions of the capture of single quanta of the magnetic flux. In this case, integer and fractional values of the plateau and steps of the Hall conductivity staircase are determined by the sizes of the sections of the edge channels containing single charge carriers.

Due to the presence of negative-U centers limiting the edge channels, macroscopic quantum phenomena are observed at different orientations of the external magnetic field when the $U_{xx}$, $U_{xy}$ voltage, and the $U_{pn}$ voltage are registered perpendicular to the QW plane studied in the framework of the Hall geometry, since in the cases under consideration the size quantization condition is satisfied.

An important role of edge channels in the detection of macroscopic quantum phenomena using galvanomagnetic techniques is manifested by varying the direction of the external magnetic field in the plane perpendicular to the SNS surface and parallel to the XY contacts from the Hall direction by an angle of 35° or 55°. In this case, both SdH oscillations and the Hall conductivity staircase are simultaneously observed, and the dominance of one or the other effect depends on the prevalence of either the Hall component of the external magnetic field or parallel to the XY contacts.

# THE AUTHORS

**Klyachkin L.E.**
e-mail: leonid.klyachkin@gmail.com
ORCID: 0000-0001-7577-1262

**Bagraev N.T.**
e-mail: bagraev@mail.ioffe.ru
ORCID: 0000-0001-8286-3472

**Malyarenko A.M.**
e-mail: annamalyarenko@mail.ru
ORCID: 0000-0002-4667-7004